\documentclass[journal]{new-aiaa}
\usepackage[utf8]{inputenc}
\usepackage{textcomp}

\usepackage{graphicx}
\usepackage{amsmath}
\usepackage[version=4]{mhchem}
\usepackage{siunitx}
\usepackage{longtable,tabularx}

\usepackage{bm}
\usepackage{xcolor}
\usepackage{siunitx}
\usepackage{cancel}
\usepackage{soul}
\usepackage{booktabs}

\title{Global stability analysis of flow behind an upswept aftbody}

\author{Qiong Liu \footnote{Postdoctoral Research Associate, Department of Mechanical and Aerospace Engineering, liu.9292@osu.edu.} and Datta V. Gaitonde\footnote{John Glenn Chair Professor, Department of Mechanical and Aerospace Engineering, 201 W 19th Ave, Columbus, AIAA Fellow.}}
\affil{Department of Mechanical and Aerospace engineering, The Ohio State University, Columbus, OH-43220, USA}
\author{Rajesh Ranjan\footnote{Assistant Professor, Department of Aerospace engineering, Indian Institute of Technology Kanpur, India, 208016.}}
\affil{Department of Aerospace Engineering, Indian Institute of Technology Kanpur, India, 208016}

\begin{document}

\newcommand{\td}{$\ang{20}$}
\newcommand{\ttd}{$\ang{32}$}
\newcommand{\ua}{upsweep angle}
\newcommand{\CR}[1] {{\color{red}{#1}}}

\maketitle

\section{Introduction}
\begin{figure*}
\centering
\includegraphics [width=\textwidth]{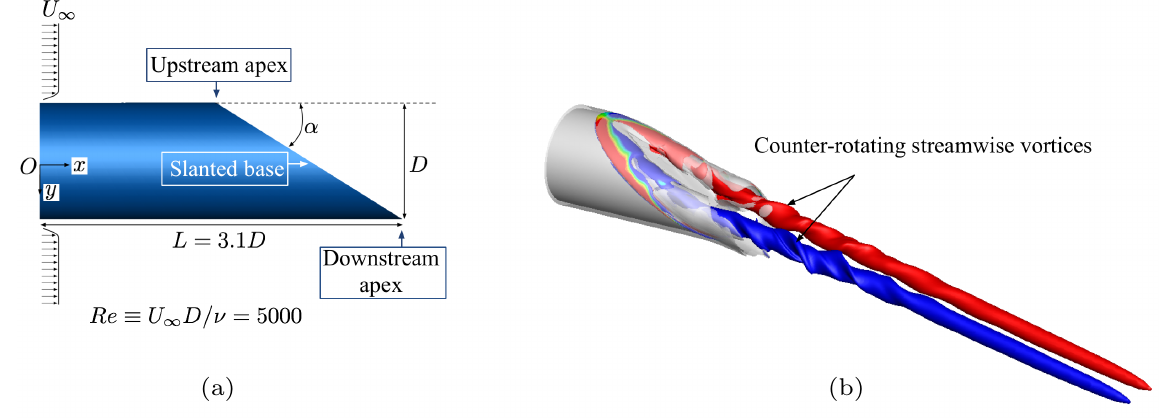}
\caption{(a) Problem description: flow over aft-body with slanted base at \ua~of $\alpha$. (b) Nature of counter-rotating streamwise vortices formed behind aftbody at flow condition of $Re=5,000$ and $\alpha=\ang{32}$. Visualized using instantaneous vorticity magnitude $||\omega||=15$ (gray) and $Q$-criterion $=10$ colored by $\omega_x$. }
\label{fig:1}	
\end{figure*}

Aft-bodies with relatively flat slanted bases are often encountered in aeronautical, automobile or maritime applications. 
Commonly, a suitable surrogate for research considers a cylinder with axis parallel to the freestream and a planar base at an angle, as illustrated in Fig.~\ref{fig:1}(a). One of the critical parameters for this geometry is upsweep angle $\alpha$. 
For a given Reynolds number, different values of  $\alpha$ generate different wake patterns with distinct flow characteristics~\cite{morel1978effect}.
Typically, the turbulent wake shedding (``wake'' regime) is observed when $\alpha \gtrsim \alpha_{cr}$, while counter-rotating streamwise vortices (``vortex'' regime) arise when $\alpha \lesssim \alpha_{cr}$ (see Fig.~\ref{fig:1}(b)), where $\alpha_{cr}$ is a critical angle, typically taken to be $\ang{45}$. 
The presence and structure of streamwise vortices can increase drag~\cite{epstein1994experimental}; this is a major consideration particularly for cargo transport aircraft, where they can disrupt operations such as airdrops~\cite{smith2013reduction} and reduce fuel efficiency. 
A better understanding of the key flow characteristics is thus of  great interest from scientific as well as engineering perspectives.

Several studies have elucidated the sequence of events that result in the vortex regime of interest in this work~\cite{morel1978effect,garmann2019high,bulathsinghala2018drag,ranjan2020mean}.
Briefly, the flow separates at the edge of the slanted base and a horseshoe-shaped vortical structure emerges.
This subsequently detaches from the base, as shown in Fig.~\ref{fig:1}(b).
The legs of the horseshoe vortex turn downstream to form the observed streamwise vortex pair, whose strength increases with $\alpha$ and results in higher drag.
Despite the known drag reduction benefits of diminishing the strength of the counter-rotating streamwise vortices, there is relatively little understanding of the underlying instabilities which could provide guidance for flow control.
As such, passive control efforts to date have considered trial and error strategies such as vertical strakes~\cite{mccluney1967drag}, vortex generators~\citep{wortman1999reduction} and flaps/spoilers~\cite{bulathsinghala2018drag,gursul2018flow} on the aft-body. 
Active flow controls~\citep{jackson2020control} explored the inception and formation of the vortices through specific energy inputs and  drag reductions of $3-11\%$ were observed depending on model incidence. 
There is currently no detailed study exploring optimal flow control for this complex flowfield. 

A detailed stability analysis of this configuration may assist in assessing the potential control parameter space and actuator placement. 
Stability analyses have been performed on vortical flows formed over wingtips for the purpose of flow control. 
Far downstream of the surface, trailing edge vortices can be represented with theoretical vortex models; for example, the Batchelor vortex is often used as a proxy in order to examine stability properties.
\citet{edstrand2018parallel} used two-dimensional (2D) stability analysis to show good agreement between the features of Batchelor vortices and trailing vortices in terms of phase speeds and growth rates. 
\citet{hein2004instability} analyzed a pair of Batchelor vortices as a model for wingtip vortices.
Using stability analysis, they report stronger modification of the instability characteristics for the vortex pair system relative to the isolated vortex. 

The flowfield of interest is different from those arising from wingtips, however.
In particular, the aftbody counter-rotating vortex pair have a common origin in the horseshoe structure on the slanted base, and evolve in relatively close proximity to each other compared to those associated with wingtips.
A holistic assessment makes imperative the consideration of both the vortices in the analysis. 
\citet{ranjan2020meandering} investigated the meandering motion of the vortices in the pair using 2D stability analysis and proper orthogonal decomposition (POD) of a cross-flow plane of data. 
Far downstream, the vortices display low-rank behavior \textit{i.e.,} they can be represented by only a few POD modes.
In the near field of the base, however, the rank order of the system is no longer low. 
Thus, while results on simpler surrogates have provided insights in the far wake regions, the dynamics of the near base are very different and fundamentally three-dimensional. 
Furthermore, passive control modifications, or actuator placement, are invariably on the surface of the body, with immediate effect on the near-base region, where velocity components in all directions are of comparable magnitude, and pressure gradients are three-dimensional and relatively large. 
The increase in shed vorticity with upsweep angle further complicates the dynamics of the near-wake region. 
Therefore, two-dimensional or far field vortex flow field analysis are insufficient to obtain a complete understanding of the flowfield.

Stability properties thus must necessarily be examined in the 3D context of the entire flow, using global stability analyses~\cite{theofilis2003advances,bagheri_global_2009,loiseau_investigation_2014} as we do here.
In contrast to the 2D assumption, the 3D global modes identified by this analysis provide comprehensive information on the presence of different intrinsic instability mechanisms. 
Following an elucidation of the stability modes, we further perform the receptivity analysis to quantify the energy amplification with respect to specific global mode structures and phase information.
The present study aims to extract information that can aid flow control strategies, since the frequency, wavelength, or actuator location could be predicated on a single or multiple instability modes.

\begin{figure}
\centering
\includegraphics [width=0.8\textwidth]{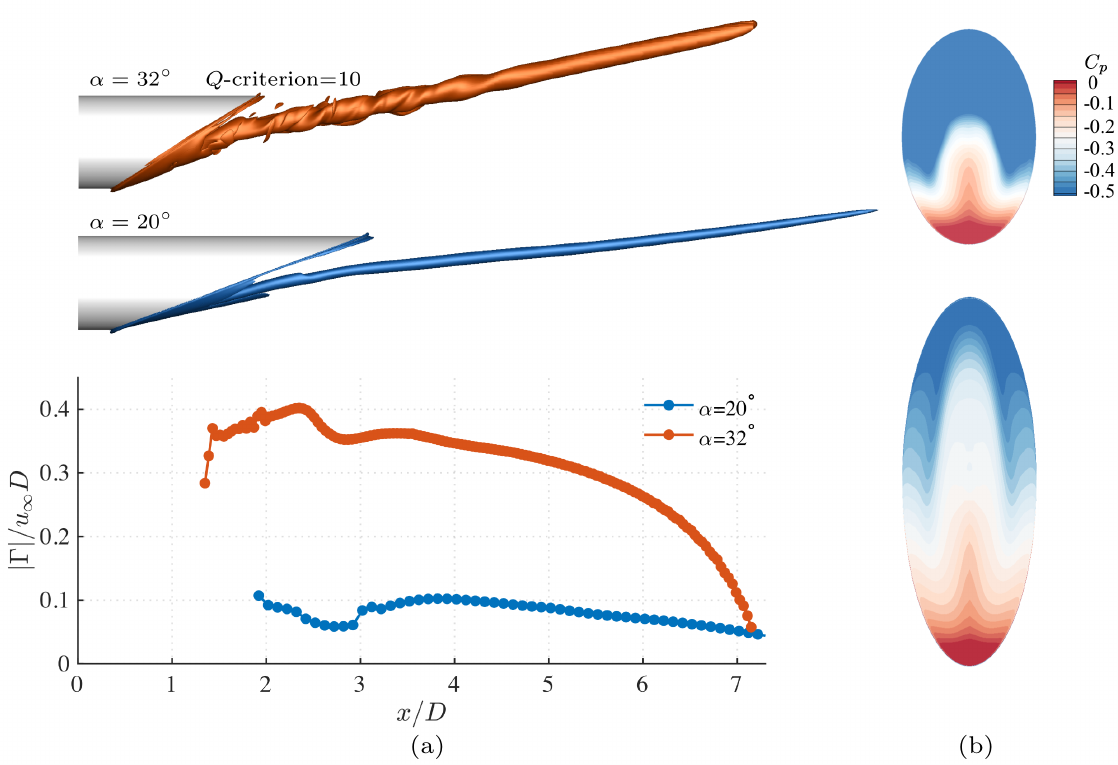}
\caption{(a) Vortical flow visualized using $Q$-criterion $=10$ and the corresponding circulation $|\Gamma|/u_\infty D$ against streamwise location. (b) Time-averaged pressure coefficient $C_p$ on the slanted base. The corresponding drag coefficients are $C_D\simeq0.3$ and $0.5$ at $\alpha=\ang{20}$ and $\ang{32}$, respectively.}
\label{fig:2}	
\end{figure}

\section{Flowfield characteristics}
We examine the flow at two different upsweep angles, $\ang{20}$ and $\ang{32}$ 
The Reynolds number, $Re=u_\infty D/\nu$, characterized by the cylinder diameter $D$ and free-stream velocity $u_\infty$ is fixed at $5{,}000$.
Although test Reynolds numbers are much higher, this low value is appropriate for the linear stability analysis being examined here~\cite{He2017JFM}: it yields a laminar, steady vortex pair at the smaller angle, and an unsteady turbulent pair at the higher angle.
All physical variables are rendered in non-dimensional form using $u_\infty$ and $D$, with pressure being normalized by $\rho u_\infty^2$. 
Large Eddy Simulations (LES) were employed to obtain the basic states. 
The numerical setup, including mesh independence and validation by comparison with experimental data, has been extensively described elsewhere~\cite{ranjan2020mean,ranjan2020meandering}.
For reference, some features of the flow, as obtained from the LES are shown in Fig.~\ref{fig:2}(a) and (b), respectively.
The top portion of Fig.~\ref{fig:2}(a) displays an iso-surface corresponding to $Q$-criterion $=10$ and serves to highlight the vertical component of the instantaneous vortex trajectories, which increase with $\alpha$.
The larger region represented by the iso-surface for $\alpha=\ang{32}$ also indicates its larger size, discussed more quantitatively below.

Two physical variables representing the effects of vortical flows are vortex strength and base pressure drag. 
The strength of the streamwise vortices is quantitatively assessed through non-dimensional circulation $\displaystyle \Gamma =\oint_C {\bf u}\cdot d{\bf l}$, where ${\bf u}$ is the velocity vector and $C$ is a closed integration path that encompasses the vorticity containing region.
The locus is chosen by varying the $Q$-criterion threshold until the results are not sensitive to this value.
The pressure drag on the slanted base is defined as $\displaystyle C_D=\frac{\int_{S} (P_\infty -P)\sin\alpha dS}{\frac{1}{2}\rho_\infty u^2_\infty A}$, where $S$ is the surface of the slanted base and $A$ is the projected area perpendicular to the free-stream direction.

The lower part of Fig.~\ref{fig:2}(a) shows the variation of $|\Gamma|/u_\infty D$ with streamwise direction for one of the vortex pairs for each case. 
Because of the highly 3D nature of the vortex near the base, the curves only start where a clear vortex can be defined -- this happens earlier for the stronger vortices in the $\ang{32}$ case.
The increase in normalized vortex strength with upsweep angle is evident; this observation agrees with the prior studies~\cite{ranjan2020meandering,zigunov2020reynolds}. 
The vortex strength displays some reduction in strength in $x/D \lesssim 3.0$ \textit{i.e.,} near the base region. 
This behavior is consistent with later stages of viscous vortex core formation, through which viscous diffusion  smears out the spiral structure to yield a smooth vorticity distribution~\cite{wu2007vorticity}. 
The corresponding pressure drag coefficients increase from $C_D\simeq0.3$ to $0.5$ as the upswept angle increases.
The pressure drag coefficient is proportional to the \ua~and is function of lower pressure footprint on the slanted base, as shown in Fig.~\ref{fig:2}(b). 
We note that when the upsweep angle exceeds the critical value, the flow changes to turbulent wake shedding and the pressure drag reduces significantly due to higher pressure recovery on the basal surface. 
This transition has been well studied and interested reader may refer to previous works~\citep{morel1978effect,ranjan2020hysteresis}.

\section{3D Stability Analysis}
To account for the complicated flowfield, we employ 3D stability analysis to explore the behavior of disturbances in the vortical flow as upsweep angle is changed and the flow bifurcates from a laminar state to a turbulent one. 
The analysis distinguishes different instabilities present in the flow and characterizes changes with \ua. 
Following standard procedure, the flow is decomposed into a steady base component $\bar{\boldsymbol u}$ and an unsteady disturbance $\epsilon{\boldsymbol u}'$, where $\epsilon$ designates a small amplitude. 
Three-dimensional, unsteady disturbance solutions are expressed in the form $\boldsymbol q'=\hat{\boldsymbol q}(x,y,z)e^{i\omega t}$, where $\hat{\boldsymbol q}(x,y,z)=[\hat{{\boldsymbol u}},\hat{p}]$ includes the velocity and pressure components and the complex number $\omega=\omega_r+\text{i}\omega_i$ contains frequency and growth rate. 
The solution of the general eigenvalue problem~\citep{peplinski2014stability,liu2016linear} yields the spatial distribution $\hat{\boldsymbol q}(x,y,z)$ of 3D instability modes with their corresponding growth ($\omega_i>0$) or damping rates ($\omega_i<0$) at circular frequencies ($\omega_r$). 
The instability results are computed with the spectral element code Nek5000~\citep{peplinski2014stability,liu2016linear}. 
 
\begin{figure*}
\centering
\includegraphics [width=0.9\textwidth]{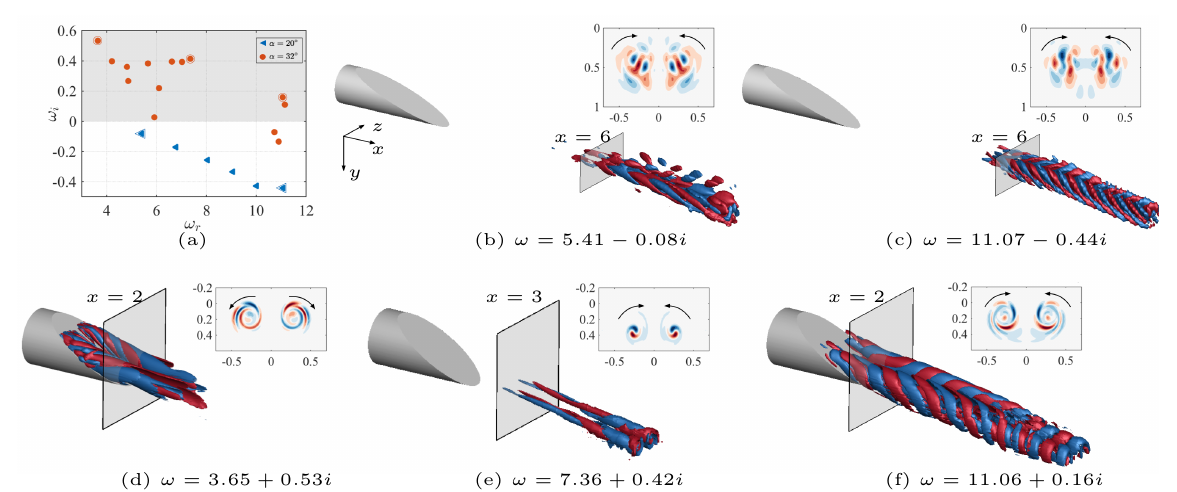}
\caption{Three-dimensional instability analysis of the vortical flows. (a) Eigenvalue spectra for cases of $\alpha=\ang{20}$ and $\ang{32}$. The unstable plane of $\omega_i>0$ are colored in grey. At $\alpha=\ang{20}$, two representative global modes are shown in (b) and (c). Global modes at $\alpha=\ang{32}$ are shown in (d),(e) and (f). The insets correspond to a cross-sectional cut at $x=6$, $2$, $3$ of real component of streamwise velocity. The solid arrows indicate the rotation directions of the modes.}
\label{fig:3}	
\end{figure*}

The variation in instability properties due to change in the upsweep angle $\alpha$ are examined first, followed by their receptivity analysis. 
The results show only stable modes  at $\alpha=\ang{20}$, while some unstable modes are obtained at $\ang{32}$.
Figure~\ref{fig:3}(a) shows eigenspectra for both cases. 
Only the positive frequency range of eigenspectra are shown due to their symmetric nature with respect to imaginary axis ($\omega_r=0$). 
The unstable modes lie in the shaded upper half-plane of real axis, where $\omega_i>0$. 
At \td, all modes lie below the neutral line $\omega_r=0$ and are thus damped. 
As the frequency is increased, the rate at which the modes are damped increases over oscillation frequencies of $5<\omega_r<12$. 
For the \ang{32} case, the flow becomes unstable (also observed in the LES) and eigenvalues emerge above the the neutral line of $\omega_r=0$. 
These unstable dynamics is assimilated into three types of modes that collectively describe the stability properties of the flowfield.

At $\ang{20}$, the stable global modes originate from the same branch and resemble vortical structures in the farfield. 
Figure~\ref{fig:3}(b) and (c) show the real components of streamwise velocity of two representative global modes. 
These have spatial structures reminiscent of vortex modes~\cite{edstrand2018parallel} but vary in their detail. 
The plot in the insert of Fig.~\ref{fig:3}(b) demonstrates the 2D structure in the $y-z$ plane at a streamwise location of $x/D=6$. 
Significant structures emerge in the downstream vortical flow field away from the aftbody and in the far-wake. 
Each global mode structure resembles  helical vortices, exhibiting symmetry along the center plane.
On account of their rotating feature, these vortical modes co-rotate with the baseline streamwise vortices, and display a gradually amplified radial structure towards the downstream. 
Due to the higher oscillation frequency, the helical mode of Fig.~\ref{fig:3}(c) exhibits smaller scales in the streamwise direction.
Such smaller scales are consistent with larger viscous diffusion effects, which reduce its growth rate. 
Consequently, this branch of vortical modes exhibits a gradual increase in the damping rate as the oscillation frequency increases (see Fig.~\ref{fig:3}(a)). 

Diverse global modes emerge at $\alpha=\ang{32}$. 
Three branches of global modes are identified by considering their oscillation frequencies and spatial structures. 
As shown in Fig.~\ref{fig:3}(a), with increasing oscillation frequency, the growth rates diminish but not in the monotonic manner of the \ang{20} case. The structures of these modes are now elaborated in more detail.

\emph{Mode I: $\omega_r+i \omega_i=3.65+0.53i$.} 
This mode pertains to the lowest oscillation frequencies among the observed global modes. 
As shown in Fig.~\ref{fig:3}(d), its structure  is  attached to the entire slanted base approaching to the upstream apex, rapidly diffusing downstream. 
Although a helical structure is observed, the sense of rotation is different from those for the \td case.
Based on an examination of the 2D structure in the $y-z$ plane at $x=2$, as shown in the insert in Fig.~\ref{fig:3}(d), the mode is revealed as antisymmetric to the center plane, which is distinct from global modes at \td.  
As such, the mode rotates in opposite manner to the baseline streamwise vortices.
This indicates that the antisymmetric instability mode is a key discriminant of the more complicated near field flow compared to \td. 
More importantly, this antisymmetric global mode should be an essential consideration for flow control, in view of its spatial distribution which includes a significant presence near the slanted surface, which are natural locations for actuator placement.

\emph{Mode II: $\omega_r+i\omega_i=7.36+0.42i$.} Mode~II exhibits a more compact radial size compared to mode~I. 
With a median range of oscillation frequency, the mode structure becomes prominent in the vortical flow away from the slanted base. 
Progressing downstream, the helical structure increases slightly in radial extent. 
The mode spirals in a co-rotational manner with the baseline streamwise vortices and possesses an azimuthal wavenumber of unity as shown in the 2D insert of Fig.~\ref{fig:3}(e). 
This vortex mode resembles the instability identified by~\citet{edstrand2018parallel}, who discovered a potential connection to vortex meandering; specifically  a helical mode of unity azimuthal wavenumber, whose kinetic energy grows monotonically in the downstream direction.

\emph{Mode~III: $\omega_r+i\omega_i=11.06+0.16i$.} 
The spatial structure of Mode~III, whose frequency is the highest among the three types of modes, 
 occupies the rear part of the slanted base and spreads rapidly downstream with dominant helical structures, as shown in Fig.~\ref{fig:3}(f). 
An examination of the 2D structure at $x/D=2$ (shown in the insert) indicates that the mode is symmetric along the center plane, a feature it shares with Mode~II. 
This mode also co-rotates with the baseline streamwise vortices. 
However, the noteworthy aspect is that mode~III has a dominant structural support that protrudes towards the rear part of the slanted base, which makes the near field vortical flow more difficult to characterize with a simple model.
This is consistent with the recent study of~\citet{ranjan2020meandering}, who showed that vortex cross-sectional data need  more POD modes for data reconstruction near the slanted base than in the far wake. 
 
Overall, the key differences from the results at $\alpha=\ang{20}$ is that, for $\ang{32}$, the emergent instability modes have a significant presence near the slanted base. 
Modes~I and~III overlap each other in a region near the slanted base. 
Since these two types of global modes display opposite rotating directions, their combined influence in the overlapping region is complicated. 
Mode~I diffuses rapidly, and thus has lower impact on the farfield. 
Therefore, perturbations introduced to mimic Mode~I have the potential to decay and may not provide significant leverage on the flow. 
Disturbances comprising Mode~III, with its prevalent structure over the aftbody and strong footprint on the vortical region,  indicates promise for a substantial interaction with the flow.

\begin{figure*}
\centering
\includegraphics [width=0.90\textwidth]{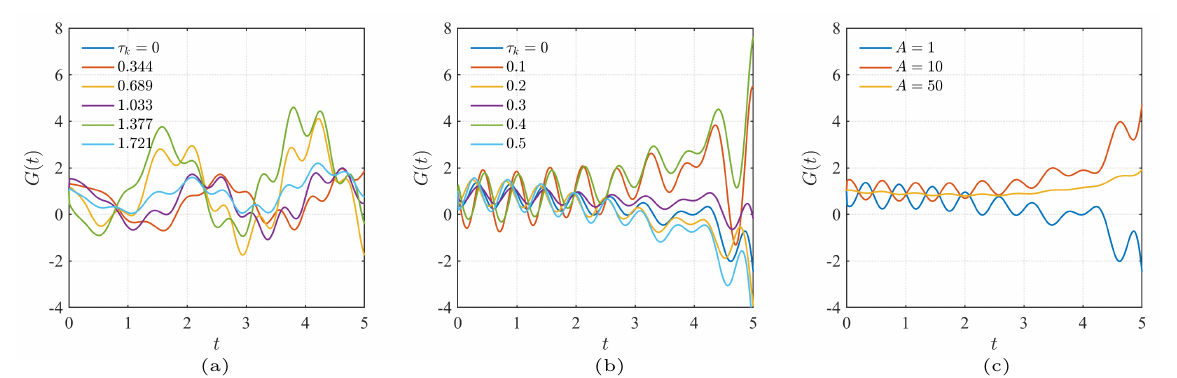}
\caption{Transfer energy $G(t)$ of (a) global mode I and (b) global mode III against time range $t\in[0,5]$ of different phases $\tau_k$. (c) Amplitude effect of transfer energy initialized by global mode III at $\tau_k=0$.}
\label{fig:4}	
\end{figure*}

\section{Receptivity Analysis}
Based on insights obtained from three-dimensional modal analysis, an assessment of the receptivity of global modes~I and~III is performed with the goal of quantifying their influence on the vortical flow from linear perspective. 
To this end, we evaluate the ratio of ouput to input energy of global modes over a time range $[0,t]$ using the amplification factor, $G(t)=\frac{E_\text{out}(t)}{E_\text{in}(0)}$.
Such an evaluation of energy transfer facilitates identification of the influence of global modes on the flowfield.
Here, the kinetic energy of perturbation is defined as $E=\bar{\boldsymbol u}{\boldsymbol u}'(t)+\frac{1}{2}{\boldsymbol u}'(t)^2$. 
The initial energy added to the linear system is prescribed through the initial perturbation. 
As measurable quantities, we set the initial condition as the real part of the global mode, \textit{i.e.,} ${\boldsymbol u}'_\text{i.c.}=C_0(re(\hat{\boldsymbol u})\cos(\omega_r \tau_k)-\text{i} \cdot im(\hat{\boldsymbol u})\sin(\omega _r\tau_k))$, where $C_0$ is the amplitude, $\omega_r$ is the circular frequency of the global mode, $re(\hat{\boldsymbol u})$ and $im(\hat{\boldsymbol u})$ are the real and imaginary parts of global modes, respectively. 
The effects of both amplitude and phase are considered. 
We select six values of $\tau_k$ which almost evenly span the entire phase space. 
The amplitudes are chosen to be $C_0=1$, $10$ and $50$. 

Figure~\ref{fig:4} shows energy amplification $G(t)$ versus time for global modes~I and~III with different phases and amplitudes. 
The flowfield exhibits different levels of energy transfer when initialized by different phases of global mode~I, especially for the phase $\tau_k=1.377$, which exhibits the largest energy peaks. 
The effect of phase becomes much clearer for the cases with global mode~III. 
Except when $\tau_k=0.1$ and $0.4$, $G(t)$ shows a reduction for the other cases over the time range $t\in[0,5]$. 
This result reveals the critical role of phase in the amplification process of energy transfer. 
Moreover, the maximum energy transfer reaches $G(t)\sim 8$ for the case initialized with global mode III.
Even though global mode~I displays a higher growth rate, the largest values of $G(t)$ are actually obtained with perturbations following mode~III. 
This highlights the manner in which global stability analyses can identify the region of flowfield sensitivity.

As an indication of the efficiency of energy transfer, we examine the effect of amplitude.
Fig.~\ref{fig:4}(c) shows the effect of amplitude on the transfer energy. 
The cases are initialized by global mode~III at phase $\tau_k=0$ and amplitude $A=1$, $10$ and $50$. 
At the smallest amplitude, $A=1$, the energy transfer reduces over time. 
The trend is reversed when the amplitude is increased to $10$, in which case, the energy gradually increases. 
A further increase to $A=50$ results in an increase over time, but at a lower magnitude than the case of $A=10$. 
One interesting observation is that the oscillatory behavior present at the other amplitudes is absent at $A=50$, which may indicate a change in the energy transfer mechanism and requires future investigation. 

\section{Summary}
The intrinsic stability modes of the 3D vortical flow behind an aft-body, characterized by a streamwise oriented counter-rotating vortex pair, are elucidated at a Reynolds number of $5{,}000$ and upsweep angles of \td and \ttd.  
The vortices are the cause of significant drag and unsteadiness, which is a concern in engineering applications. 
Although the basic state bears some similarity between the  \td~and \ttd cases, their stability modes are very distinct. 
At \td, the flow is laminar and the global modes are stable with principal structures occurring in the downstream vortical flow field, away from the aft-body.
At \ttd, three different types of global modes are classified based on the range of frequency and spatial structures. 
The lowest frequency global mode~I is antisymmetric and is attached to the entire slanted base  near the upstream apex. 
The structure diffuses relatively quick as it develops downstream. 
The symmetric high-frequency global mode~III occupies the rear part of the slanted base and is comprised of relatively small radial helical structures occurring away from the aft body.
Since both global modes~I and~III, are attached on the slanted base, a further evaluation of the energy amplification is performed to examine the effect of perturbation phase and amplitude. 
The analysis indicates that mode~III, which is prominent on the slanted base region as well as the vortical field, has a larger influence on the flow. 
The classification of the intrinsic instability properties of the vortical flow over the aft-body thus provides indispensable information for potential flow control techniques. 
In addition to location, such results aid in assessing the effect of different combinations of frequency, wavelength, or locations to excite or dampen instabilities.

\section*{Acknowledgments}
This research was partially supported by Air Force Office of Scientific Research ( FA9550-17-1-0228, Program Officer: Gregg Abate). The simulations were performed at the DoD HPCMP DSRCs.

\bibliography{ref}

\end{document}